\documentclass[11pt]{article}

\usepackage[final]{acl}

\usepackage{times}
\usepackage{latexsym}

\usepackage[T1]{fontenc}

\usepackage[utf8]{inputenc}

\usepackage{microtype}

\usepackage{inconsolata}

\usepackage{graphicx}

\usepackage{xcolor}
\definecolor{AcademicBlue}{RGB}{30, 70, 140} 

\usepackage{xspace}
\newcommand{\sys}{\textbf{\texttt{DPC}}\xspace}
\newcommand{\slicer}{\textsc{Slicer}\xspace}
\newcommand{\tester}{\textsc{Tester}\xspace}
\newcommand{\solver}{\textsc{Solver}\xspace}

\usepackage{amsmath}

\usepackage{amssymb}

\usepackage{algorithm}
\usepackage{algorithmic}

\usepackage[table]{xcolor}
\usepackage{booktabs}
\definecolor{iceblue}{rgb}{0.9, 0.95, 1.0}

\usepackage{colortbl}
\usepackage{multirow}
\usepackage{multicol}

\usepackage{enumitem}

\usepackage{colortbl} 
\usepackage{booktabs} 
\usepackage{subcaption} 
\usepackage[most]{tcolorbox}
\usepackage{listings} 
\usepackage{xcolor}

\definecolor{bg_gray}{RGB}{248, 248, 248}
\definecolor{frame_gray}{RGB}{180, 180, 180}
\definecolor{var_blue}{RGB}{0, 80, 200}

\definecolor{color_champ}{RGB}{255, 230, 230} 
\definecolor{color_chal}{RGB}{230, 255, 230} 
\definecolor{code_keyword}{RGB}{0, 0, 255}
\definecolor{code_comment}{RGB}{0, 128, 0}

\newcommand{\pvar}[1]{\textcolor{var_blue}{\textbf{#1}}}

\newtcolorbox{promptframe}[1]{
  enhanced,
  breakable,             
  title=\textbf{#1},     
  colframe=frame_gray,   
  colback=bg_gray,       
  coltitle=white,        
  colbacktitle=black!70, 
  arc=2mm,               
  boxrule=0.8pt,         
  top=0mm, bottom=0mm, left=1mm, right=1mm, 
}

\lstdefinestyle{promptstyle}{
  basicstyle=\scriptsize\ttfamily, 
  breaklines=true,                 
  columns=fixed,                   
  keepspaces=true,                 
  escapeinside={(*@}{@*)},         
  aboveskip=2pt,
  belowskip=2pt,
  showstringspaces=false,
  extendedchars=true
}

\usepackage[utf8]{inputenc}
\usepackage[T1]{fontenc}
\usepackage{geometry}
\usepackage{xcolor}
\usepackage[most]{tcolorbox}
\usepackage{fontawesome5} 

\definecolor{bg_gold}{RGB}{245, 245, 245}
\definecolor{bg_sc}{RGB}{255, 245, 240}
\definecolor{frame_sc}{RGB}{200, 50, 50}
\definecolor{bg_dpc}{RGB}{240, 255, 240}
\definecolor{frame_dpc}{RGB}{34, 139, 34}

\tcbset{
    colback=white,
    colframe=black!70,
    boxrule=0.5mm,
    arc=3mm,
    fonttitle=\bfseries,
    enhanced,
    drop shadow
}

\title{{\color{AcademicBlue}\texttt{DPC}}: Training-Free Text-to-SQL Candidate Selection via {\color{AcademicBlue}D}ual-{\color{AcademicBlue}P}aradigm {\color{AcademicBlue}C}onsistency}

\author{
Boyan Li,
Ou Ocean Kun Hei,
Yue Yu,
Yuyu Luo\textsuperscript{\textdagger} \\
The Hong Kong University of Science and Technology (Guangzhou) \\
\textsuperscript{\textdagger}Corresponding author \\
\small{\textbf{Code:} \url{https://github.com/HKUSTDial/DPC}}
}

\begin{document}

\setlength{\abovedisplayskip}{3pt plus 1pt minus 1pt}
\setlength{\belowdisplayskip}{3pt plus 1pt minus 1pt}
\setlength{\abovedisplayshortskip}{2pt plus 1pt minus 1pt}
\setlength{\belowdisplayshortskip}{2pt plus 1pt minus 1pt}

\maketitle

\begin{abstract}
While Large Language Models (LLMs) demonstrate impressive proficiency in generating SQL queries, they fundamentally lack the capability to self-evaluate correctness without an execution oracle. 
This limitation creates a stark \textbf{\textit{Generation-Selection Gap}}, where high potential accuracy (\textit{Pass@K}) fails to translate into execution accuracy (\textit{Pass@1}).
Although supervised verifiers offer mitigation, they incur prohibitive annotation costs and suffer from domain fragility. 
Consequently, recent research has pivoted to the \textit{training-free} setting. 
However, existing methods---such as Self-Consistency or LLM-as-a-Judge---remain hampered by \textit{systematic bias} (consensus on hallucinations) and \textit{symbolic blindness} (inability to simulate execution states). 
We introduce \sys (Dual-Paradigm Consistency), a multi-agent framework that reformulates SQL selection from a probabilistic guessing task on hidden data into a deterministic verification task on visible data. 
Specifically, \sys employs a \slicer and a \tester agent to collaboratively construct a Minimal Distinguishing Database (MDD)---an adversarial, fully observable micro-environment engineered to expose logical discrepancies between candidates. 
To break the self-correction bias, a \solver agent then verifies the SQL candidates by cross-referencing their execution against a parallel Python/Pandas solution. 
By validating \textit{execution consistency} between declarative (SQL) and imperative (Python) paradigms, \sys robustly discriminates correct logic from systematic hallucinations. 
Experiments on BIRD and Spider across multiple LLMs demonstrate that our method consistently outperforms existing selection baselines, achieving absolute accuracy improvements of up to 2.2\% over strong competitors like Self-Consistency.
\end{abstract}

\section{Introduction}

Recent advances in Large Language Models (LLMs) have revolutionized the Text-to-SQL task, enabling systems to generate highly plausible SQL queries for complex questions~\cite{dataset-bird, dataset-spider, elliesql, nl2sql360}. However, due to the inherent stochasticity of LLMs, a single generation is often unreliable. Consequently, state-of-the-art approaches adopt a ``\textbf{\textit{generate-then-select}}'' paradigm: generating $K$ candidate queries to cover the correct logic~\cite{alpha-sql, deepeye-sql}. In this work, we focus on the critical downstream task: \textbf{SQL Candidate Selection}---the process of autonomously identifying the single correct executable query from a pool of $K$ candidates.

\begin{figure}[t]
    \centering
    \includegraphics[width=\linewidth]{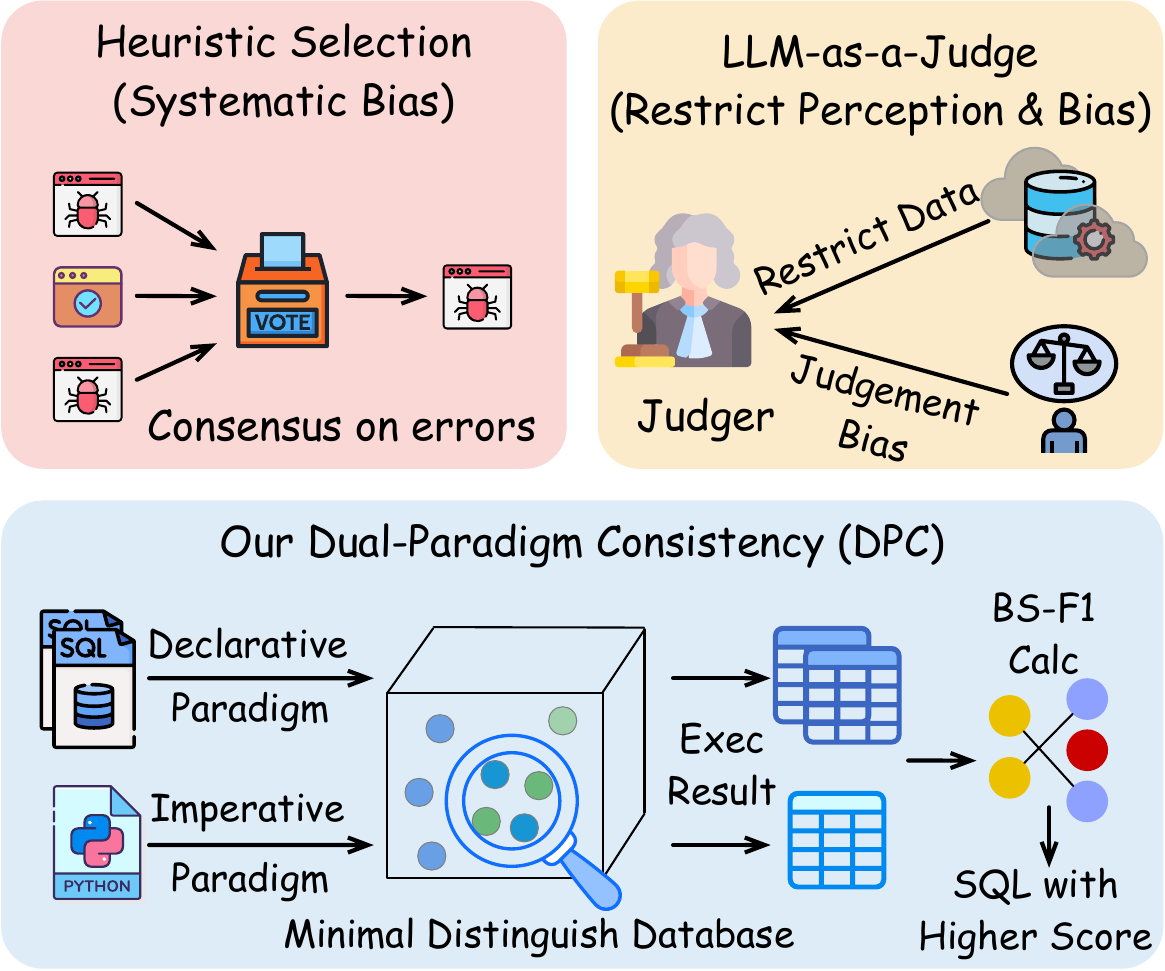}
\caption{
    \textbf{Comparison of Selection Paradigms.} 
    (Top-Left) \textit{Heuristic Selection} relies on majority voting but fails when models exhibit \textbf{systematic bias}, reaching a consensus on errors.
    (Top-Right) \textit{LLM-as-a-Judge} suffers from \textbf{restricted perception} due to the lack of execution feedback, relying solely on internal priors.
    (Bottom) Our \sys framework introduces a \textit{Dual-Paradigm} approach (SQL \& Python). It synthesizes a \textbf{Minimal Distinguishing Database} to execute both paradigms, determining correctness through deterministic result consistency rather than probability.
}
    \label{fig:comparison}
    \vspace{-1em}
\end{figure}

Despite remarkable generative capabilities, a significant \textbf{\textit{Generation-Selection Gap}} persists.
As shown in Table~\ref{tab:selection_comparison}, while the potential \textit{Pass@K} is high (e.g., 58.8\% on BIRD), \textbf{\textit{real-world users demand a single correct answer}}; yet, the actual \textit{Pass@1} significantly lags behind ($\sim$50\%).
Although \textit{training-based verifiers}~\cite{xiyan-sql} attempt to bridge this via fine-tuning, prohibitive annotation costs and poor generalization limit their utility.
Consequently, focus has shifted to the \textbf{training-free} setting, yet current mechanisms remain inadequate.
\textit{Heuristic approaches} like Self-Consistency~\cite{opensearch-sql} fail under \textit{systematic bias}, where models consistently converge on errors (Figure~\ref{fig:comparison}, Top-Left).
Similarly, \textit{LLM-as-a-Judge}~\cite{mcs-sql} is hindered by \textit{symbolic blindness}: constrained by limited data context and unable to mentally simulate execution, it relies on internal priors and exhibits \textit{intrinsic judgement bias} (Figure~\ref{fig:comparison}, Top-Right).

We argue that the failure of current selection methods stems from three intrinsic challenges in the Text-to-SQL paradigm. First, \textbf{(C1) Partial Observability}: Real-world databases are typically too massive to fit within the context window. This forces the model to verify logic against a ``hidden'' data distribution ($\mathcal{D}_{hidden}$) based on probabilistic assumptions rather than concrete evidence, creating an epistemic gap that precludes rigorous verification. Second, \textbf{(C2) Symbolic Blindness}: Even if the data were visible, LLMs lack an internal interpreter to reliably simulate the state changes of complex SQL operations (e.g., nested JOINs) merely by inspection. Finally, \textbf{(C3) Intrinsic Confirmation Bias}: When acting as a selector, the model inherently favors candidates that align with its internal priors---even if flawed. This results in a ``conflict of interest'' where the model fails to objectively distinguish between its own hallucinations and correct logic.

To address these challenges, we propose \sys (Dual-Paradigm Consistency), a multi-agent framework that shifts SQL selection from a probabilistic guessing task on hidden data to a \textit{deterministic reasoning task on visible data}.
\sys follows a progressive verification pipeline. First, to overcome \textbf{Partial Observability (C1)}, we introduce the concept of \textit{Adversarial Environment Synthesis}. Instead of reasoning about the massive, invisible database, \sys employs a \slicer and a \tester agent to construct a \textbf{Minimal Distinguishing Database (MDD)}. The MDD is not merely a data sample, but a context-fitting, adversarial micro-environment specifically engineered to yield divergent execution results for conflicting SQLs. This transforms the verification environment from partially observable to \textbf{Fully Observable}, enabling the model to ground its decisions in explicit execution evidence.

Second, to mitigate \textbf{Symbolic Blindness (C2)} and \textbf{Intrinsic Confirmation Bias (C3)}, we introduce \textbf{Dual-Paradigm Verification}. 
Rather than relying solely on SQL generation, we leverage the model's inherent \textit{competency disparity} across programming languages. Extensive research indicates that LLMs exhibit superior reasoning capabilities in widespread imperative languages like Python, attributed to their dominance in pre-training corpora relative to domain-specific query languages~\cite{DBLP:journals/corr/abs-2503-17181, DBLP:journals/corr/abs-2402-19173}. 
Furthermore, translating a user's intent into imperative Python code forces the model to explicitly plan the data manipulation steps, offering a distinct reasoning path from declarative SQL formulation~\cite{DBLP:conf/icml/GaoMZ00YCN23, DBLP:journals/corr/abs-2502-19411}.
Therefore, \sys employs a \solver agent to generate a parallel Python solution on the MDD. By treating the higher-confidence Python solution as a \textit{proxy ground truth}, \sys pinpoints the correct SQL candidate that aligns with the imperative logic.

Our contributions are summarized as follows:
\begin{itemize}[leftmargin=*, noitemsep, topsep=0pt, parsep=0pt]
    \item We identify \textit{Partial Observability} and \textit{Systematic Bias} as the core constraints in SQL selection and propose the \textbf{Minimal Distinguishing Database (MDD)} to facilitate rigorous, fully observable, and adversarial verification.
    \item We introduce \sys, a training-free framework that leverages \textbf{Dual-Paradigm Consistency} to identify semantic correctness via complementary reasoning paths.
    \item We design a \textbf{Bipartite Soft-F1 (BS-F1)} metric to handle formatting heterogeneities and row-ordering ambiguity between SQL and Python results, ensuring rigorous alignment.
    \item Extensive experiments on BIRD and Spider benchmarks demonstrate that \sys consistently outperforms state-of-the-art training-free selection methods, establishing new standards in execution accuracy.
\end{itemize}
\section{Problem Formulation}

\subsection{The SQL Selection Task}
Let $Q$ denote a natural language question and $S = (\mathcal{T}, \mathcal{C}, \mathcal{R})$ denote a relational database schema, where $\mathcal{T}$ is a set of tables, $\mathcal{C}$ is a set of columns, and $\mathcal{R}$ represents foreign key relations.
Given $Q$ and $S$, an LLM generates a set of $K$ candidate SQL queries, denoted as $\mathcal{Y} = \{y_1, y_2, \dots, y_K\}$.

The objective of SQL Selection is to identify the optimal candidate $y^* \in \mathcal{Y}$ that is semantically equivalent to the ground truth query $y_{gt}$. Semantic equivalence is defined by execution correctness on the database instance:
\begin{equation}
    \textsc{Exec}(y^*, \mathcal{D}) = \textsc{Exec}(y_{gt}, \mathcal{D})
\end{equation}
where $\mathcal{D} = (S, \mathcal{V})$ represents the database instance, consisting of the schema $S$ and the actual data values (records) $\mathcal{V}$.

In the \textbf{training-free} setting, we seek a selection function $f$ that maximizes the likelihood of identifying $y^*$ under fixed model parameters $\theta$:
\begin{equation}
    f^* = \operatorname*{argmax}_{f} P\big(f(Q, S, \mathcal{Y}; \theta) \equiv y_{gt}\big)
\end{equation}
subject to the constraint that $f$ operates without access to the ground truth $y_{gt}$ or gradient updates.

\subsection{The Observability Gap}
A critical constraint in real-world deployment is that the data instance is \textit{partially observable}.
Let $W$ denote the context window capacity of the LLM. Typically, the full data values $\mathcal{V}$ far exceed this limit ($|\mathcal{V}| \gg W$).
While standard prompting strategies may provide a small set of sample values $\mathcal{V}_{example} \subset \mathcal{V}$ (e.g., 3-shot row samples) to the model, the vast majority of the data distribution remains unseen, denoted as $\mathcal{V}_{unseen} = \mathcal{V} \setminus \mathcal{V}_{example}$.

This partial observability creates a \textit{verification gap}. The provided $\mathcal{V}_{example}$ is typically static and sparse, lacking the \textit{boundary cases} necessary to verify complex logic. For instance, distinguishing between \texttt{INNER JOIN} and \texttt{LEFT JOIN} requires specific records (e.g., unmatched keys) that are likely absent in a random sample.
Consequently, execution consistency on the example subset is a necessary but insufficient condition for correctness:
\begin{equation}
    \resizebox{0.89\linewidth}{!}{%
        $ 
        \begin{split}
            \textsc{Exec}(y_i, \mathcal{D}_{example}) &= \textsc{Exec}(y_{gt}, \mathcal{D}_{example}) \\
            &\not\Rightarrow y_i \equiv y_{gt}
        \end{split}
        $ %
    }
\end{equation}
where $\mathcal{D}_{example} = (S, \mathcal{V}_{example})$. The model is thus forced to select $y^*$ based on probabilistic priors rather than definitive execution evidence.


\subsection{Objective Reformulation}
\label{sec:objective_reformulation}
Since verification on partial $\mathcal{D}_{example}$ is unreliable, we reformulate the selection task by constructing a \textbf{Minimal Distinguishing Database (MDD)}, denoted as $\mathcal{D}_{MDD} = (S', \mathcal{V}_{MDD})$. To serve as a valid verification ground, the MDD must satisfy two core properties: \textbf{(i) Contextual Feasibility}, requiring the synthetic instance to fit within the LLM's context window ($|\mathcal{D}_{MDD}| \leq W$); and \textbf{(ii) Discriminative Validity}, where $\mathcal{V}_{MDD}$ is adversarially engineered to expose semantic discrepancies. Formally, for any pair of semantically distinct candidates $y_i, y_j \in \mathcal{Y}$ (where $y_i \not\equiv y_j$), their execution on the MDD must diverge:
\begin{equation}
    \resizebox{0.89\linewidth}{!}{%
        $ 
        y_i \not\equiv y_j \implies \textsc{Exec}(y_i, \mathcal{D}_{MDD}) \neq \textsc{Exec}(y_j, \mathcal{D}_{MDD})
        $ %
    }
\end{equation}
Under these constraints, \sys transforms the task from probabilistic guessing on hidden data to deterministic consistency checking on visible data.
\section{The \sys Framework}

\begin{figure*}[t]
    \centering
    \includegraphics[width=1.0\textwidth]{figs/dpc_overview.pdf} 
    \caption{
        \textbf{Overview of the \sys framework.} 
        The pipeline transforms selection into verification by identifying conflicting candidates (Champion vs. Challenger) and synthesizing an adversarial \textbf{Minimal Distinguishing Database (MDD)}. 
        By executing candidates alongside a generated Python reference on this fully observable environment, \sys utilizes the \textbf{BS-F1} to deterministically identify the correct logic via cross-paradigm consistency.
    }
    \label{fig:framework}
    \vspace{-1em}
\end{figure*}

\subsection{Framework Overview}
\sys is a multi-agent system bridging the epistemic gap in SQL selection. 
As shown in Figure \ref{fig:framework}, it reformulates the task into a \textit{progressive verification pipeline}, prioritizing external execution evidence over internal priors. 
It coordinates three agents (\slicer, \tester, \solver) across four phases (see Appendix~\ref{sec:appendix_prompts} for detailed prompts):

\begin{description}[style=unboxed, leftmargin=0cm, itemsep=2pt, parsep=0pt, topsep=2pt]
    \item[\textbf{Stage 1: Candidate Clustering \& Pairing.}] 
    Instead of evaluating all candidates individually, \sys first condenses the selection space by clustering semantically identical queries. It then samples a \textit{Champion-Challenger} pair to represent the most significant logical conflict, maximizing the efficiency of subsequent verification.
    
    \item[\textbf{Stage 2: Adversarial Environment Synthesis.}] 
    To resolve the conflict, the \slicer and \tester agents collaborate to synthesize the \textbf{Minimal Distinguishing Database (MDD)}. Unlike random sampling, the MDD is adversarially generated to be \textit{context-fitting} (small enough for the context window) yet \textit{discriminative} (guaranteeing divergent execution results for conflicting logic).

    \item[\textbf{Stage 3: Dual-Paradigm Execution.}] 
    With the environment fully observable, \sys introduces a complementary reasoning path. The \solver agent generates an imperative Python script to solve the question on the MDD. This Python execution serves as a high-confidence \textit{reference anchor} for verifying the declarative SQL candidates.

    \item[\textbf{Stage 4: Consistency Verification.}] 
    Finally, \sys determines the winner by comparing the SQL execution results against the Python reference. We employ a novel \textbf{Bipartite Soft-F1} metric to robustly quantify semantic equivalence, overcoming both formatting heterogeneities and row-ordering ambiguity inherent to cross-paradigm verification.
\end{description}

\subsection{Stage 1: Candidates Clustering \& Pairing}
\label{sec:stage1}
The primary objective of this stage is to distill the noisy candidate set into two representative SQLs for verification. Since LLMs generate SQL queries probabilistically, the raw output often contains syntactic variations that are logically equivalent (e.g., varying alias names or keyword capitalization).

\paragraph{Execution-Consistency Clustering.} 
We adopt an execution-based strategy to simplify the candidate set. Let $\mathcal{D}$ denote the original database instance available for execution. We execute every candidate $y_{k}\in\mathcal{Y}$ on $\mathcal{D}$ to obtain its execution result $E_{k}$. Candidates are then grouped into clusters $\mathcal{C}_{1},\mathcal{C}_{2},...,\mathcal{C}_{m}$ such that all queries within a cluster produce identical execution results:
\begin{equation}
    \resizebox{0.89\linewidth}{!}{%
        $y_a, y_b \in \mathcal{C}_i \iff \textsc{Exec}(y_a, \mathcal{D}) = \textsc{Exec}(y_b, \mathcal{D})$%
    }
\end{equation}

\paragraph{Champion-Challenger Selection.} 
To maximize the efficiency of the subsequent verification, we select the two most dominant SQLs for a pairwise duel~\cite{deepeye-sql, csc-sql}. We sort the clusters by size $|\mathcal{C}_i|$ in descending order to identify two candidates: the \textbf{Champion} ($y_{champ}$), selected from the largest cluster $\mathcal{C}_{max}$, which represents the model's ``Majority Vote'' choice and serves as a strong baseline; and the \textbf{Challenger} ($y_{chal}$), from the second-largest cluster, representing the most probable alternative hypothesis that conflicts with the consensus.

\subsection{Stage 2: Adversarial Environment Synthesis}
\label{sec:stage2}

To resolve the conflict between the Champion ($y_{champ}$) and the Challenger ($y_{chal}$), \sys constructs a \textbf{Minimal Distinguishing Database (MDD)}. This synthetic environment is engineered to satisfy the contextual feasibility and discriminative validity defined in Section~\ref{sec:objective_reformulation}. See Appendix~\ref{sec:appendix_running_example} for a detailed running example.

\paragraph{The \slicer Agent.}
The full database schema $S$ often contains numerous tables and columns irrelevant to the current query, introducing noise and consuming context window space. The \slicer Agent iteratively distills a focused schema subgraph $S' \subseteq S$ that contains only the tables and columns necessary for the candidate SQLs.

To ensure the structural integrity of the sliced schema (e.g., ensuring no referenced columns or foreign keys are missing), we implement a \textit{Dry-Run Validation} loop. In each iteration $t$, the agent predicts a schema candidate $S'_t$. We then attempt to execute (or \texttt{EXPLAIN}) both $y_{champ}$ and $y_{chal}$ on an empty database instance structured with $S'_t$. If execution fails, the database error message $e_t$ is fed back to the agent for self-correction:
\begin{equation}
    S'_{t+1} = \text{\slicer}(Q, S, \{y_{champ}, y_{chal}\}, e_t)
\end{equation}
The process terminates upon dry-run success or reaching the maximum iteration limit $T_{max}$.

\paragraph{The \tester Agent.}
Based on the validated slice $S'$, the \tester Agent constructs the synthetic data values $\mathcal{V}_{MDD}$ to populate the MDD. Unlike random data generation, this process is adversarial: the goal is to generate specific records that expose the semantic discrepancy between the two candidates.

We employ a \textit{Discriminative Feedback} loop to guarantee effectiveness. In iteration $t$, the agent generates a data sample $\mathcal{V}_t$. We execute both candidates on $\mathcal{D}_t = (S', \mathcal{V}_t)$ and compare their outputs. If the outputs are identical, it implies the data is insufficient to distinguish the logic (e.g., lacking a specific boundary case for a \texttt{JOIN}). The agent receives a penalty signal and regenerates the data:
\begin{equation}
\mathcal{V}_{t+1} = \text{\tester}(Q, S', y_{champ}, y_{chal}, \mathbb{I}_{equal})
\end{equation}
where $\mathbb{I}_{equal}$ is the feedback indicator. The loop continues until the condition is met or the maximum iteration limit $T_{max}$ is reached:
\begin{equation}
\resizebox{0.89\linewidth}{!}{%
    $\textsc{Exec}(y_{champ}, \mathcal{D}_{MDD}) \neq \textsc{Exec}(y_{chal}, \mathcal{D}_{MDD})$%
}
\end{equation}

\subsection{Stage 3: Dual-Paradigm Execution}
\label{sec:stage3}

With the fully observable MDD established, \sys performs dual-paradigm execution. We execute the candidate SQLs ($y_{champ}$ and $y_{chal}$) on the $\mathcal{D}_{MDD}$ to obtain their deterministic outputs, denoted as $E_{champ}$ and $E_{chal}$ respectively. Simultaneously, to verify these results against an independent reasoning path, this stage employs a \solver Agent to solve the user question $Q$ using Python (Pandas).

\paragraph{The Solver Agent.}
To establish a reliable verification standard, we leverage the model's \textit{competency disparity} between paradigms. Imperative Python code benefits from superior pre-training coverage~\cite{DBLP:journals/corr/abs-2503-17181, DBLP:journals/corr/abs-2402-19173} and explicit step-by-step logic derivation~\cite{DBLP:conf/icml/GaoMZ00YCN23, DBLP:journals/corr/abs-2502-19411}, offering a higher-confidence reasoning path than declarative SQL. Consequently, we treat the Python execution result as a robust \textit{reference anchor} (proxy ground truth) to validate the SQL candidates.

To guarantee the executability of the generated logic, we implement a \textit{Runtime Self-Correction} loop. The agent initially generates a Python script $\rho_0$ based on the schema and data in $\mathcal{D}_{MDD}$. In each iteration $t$, we execute $\rho_t$ within a sandboxed Python environment. If the execution raises an exception (e.g., \texttt{KeyError}, \texttt{SyntaxError}), the traceback message $e_t$ is captured and fed back to the agent to guide the debugging process:
\begin{equation}
    \rho_{t+1} = \solver(Q, \mathcal{D}_{MDD}, \rho_t, e_t)
\end{equation}
This iterative refinement continues until the script executes successfully or the retry limit $T_{max}$ is reached, yielding the final execution result $E_{py}$.

\subsection{Stage 4: Consistency Verification}
\label{sec:stage4}

\begin{algorithm}[t!]
\small
\caption{\small Bipartite Soft-F1 (BS-F1) Calculation}
\label{alg:bipartite_f1}

\renewcommand{\algorithmiccomment}[1]{\hfill $\triangleright$ \textit{#1}}

\begin{algorithmic}[1]
\renewcommand{\algorithmicrequire}{\textbf{Input:}}
\renewcommand{\algorithmicensure}{\textbf{Output:}}

\REQUIRE SQL Result $E_{sql}$, Python Proxy $E_{py}$
\ENSURE Consistency Score $s \in [0, 1]$

\STATE \textbf{Preprocessing:}
\STATE $R_{sql} \gets \textsc{Normalize}(E_{sql})$
\STATE $R_{py} \gets \textsc{Normalize}(E_{py})$

\STATE \textbf{Global Optimal Matching:}
\STATE Let $N = |R_{sql}|$, $M = |R_{py}|$
\STATE Initialize Cost Matrix $C$ of size $N \times M$
\FOR{$i \in 0 \dots N-1, \ j \in 0 \dots M-1$}
    \STATE $m_{ij} \gets$ column overlap ratio between $R_{sql}[i]$ and $R_{py}[j]$
    \STATE $C_{i,j} \gets 1.0 - m_{ij}$ \COMMENT{Minimize cost $\equiv$ Maximize overlap}
\ENDFOR

\STATE \textit{$\triangleright$ Solve Assignment Problem via Hungarian Algorithm}
\STATE $\mathcal{A} \gets \textsc{HungarianAlgorithm}(C)$ 
\STATE Initialize $TP, FP, FN \gets 0, 0, 0$

\STATE \textbf{Aggregation:}
\FOR{assigned pair $(i, j)$ in $\mathcal{A}$}
    \STATE Accumulate $TP, FP, FN$ based on match of $(R_{sql}[i], R_{py}[j])$
\ENDFOR

\STATE \textit{$\triangleright$ Handle Unmatched Rows (Penalties)}
\STATE $FP \mathrel{+}= (N - |\mathcal{A}|)$ \COMMENT{Unmatched SQL rows}
\STATE $FN \mathrel{+}= (M - |\mathcal{A}|)$ \COMMENT{Unmatched Python rows}

\STATE \textbf{Final Calculation:}
\STATE $P \gets \frac{TP}{TP + FP}, \quad R \gets \frac{TP}{TP + FN}$
\RETURN $2 \cdot \frac{P \cdot R}{P + R}$
\end{algorithmic}
\end{algorithm}

The final stage is to determine which candidate---the Champion or the Challenger---aligns with the verified Python reference anchor $E_{py}$.

\paragraph{The Challenge of Heterogeneity.}
In standard Text-to-SQL evaluation, \textbf{Execution Accuracy (EX)} checks for strict identity between result sets. However, this rigid metric fails in our cross-paradigm setting ($E_{sql}$ vs. $E_{py}$) due to:
\textbf{(i) Type Incompatibility:} SQL and Python/Pandas utilize different internal representations for semantically identical values (e.g., SQL \texttt{DECIMAL} vs. Python \texttt{float}, SQL \texttt{NULL} vs. Python \texttt{NaN}), causing standard EX to reject valid matches.
\textbf{(ii) Ordering Ambiguity:} SQL query results without \texttt{ORDER BY} are unordered sets, whereas Pandas operations often produce implicitly ordered indices, leading to false negatives despite identical content.

To bridge this gap without introducing human intervention, we propose the \textbf{Bipartite Soft-F1 (BS-F1)}, a robust deterministic metric grounded in combinatorial optimization.

\paragraph{BS-F1 Calculation.} As detailed in Algorithm~\ref{alg:bipartite_f1}, the BS-F1 metric addresses heterogeneity through a three-step pipeline: \textbf{(i) Atomic Value Normalization:} We apply type-agnostic mapping (Table~\ref{tab:norm_rules}) to unify diverse data types into a canonical format, resolving type incompatibilities. \textbf{(ii) Global Optimal Matching:} To handle ordering ambiguity, we model result alignment as a \textit{Maximum Weight Bipartite Matching} problem. By constructing a cost matrix of semantic distances between SQL and Python rows, we utilize the \textbf{Hungarian Algorithm}~\cite{kuhn-algo, munkres-algo, hungarian-algo} to find the global optimal assignment maximizing column-level overlap. \textbf{(iii) Score Aggregation:} Based on the optimal assignment, we aggregate True Positives (TP) while unmatched rows (extra or missing data) strictly penalize the score as FP or FN. The harmonic mean of the resulting Precision and Recall yields the final BS-F1 score.

\begin{table}[t]
\centering
\small
\caption{Atomic Value Normalization used in \sys.}
\label{tab:norm_rules}
\vspace{-1em}
\begin{tabular}{l l}
\toprule
\textbf{Original Type/Value} & \textbf{Normalized Target} \\
\midrule
\texttt{Decimal}, \texttt{float} & \texttt{float} (rounded to 4 decimals) \\
\texttt{datetime}, \texttt{Timestamp} & ISO String (\texttt{YYYY-MM-DD}) \\
\texttt{None}, \texttt{NaN}, "null" & \texttt{None} (Null Object) \\
String w/ whitespace & Stripped String \\
\bottomrule
\end{tabular}
\end{table}

The candidate with the highest BS-F1 score is selected as the final prediction $y^*$:
\begin{equation}
y^* = \operatorname*{argmax}_{y \in \{y_{champ}, y_{chal}\}} \text{BS-F1}(E_{y}, E_{py})
\end{equation}
By grounding verification in a metric that is robust to format heterogeneity yet sensitive to semantic logic, \sys effectively mitigates the intrinsic bias of LLM self-correction.

\section{Experiments}
\label{sec:exp}

We conduct comprehensive experiments to verify the effectiveness and efficiency of \sys, benchmarking it against both general-purpose base LLMs and state-of-the-art NL2SQL specialized solutions.

\begin{table*}[t!]
\centering
\small
\caption{Comparison of different SQL selection methods using multiple base LLMs on BIRD and Spider datasets. \textbf{Upper Bound (Pass@N)} indicates the theoretical maximum accuracy within the candidate pool ($N=5$). The best results among selection methods are \textbf{bolded}, and the second-best results are \underline{underlined}.}
\label{tab:selection_comparison}
\setlength{\extrarowheight}{1.5pt} 
\vspace{-1em}
\begin{tabular}{l ccc c ccc}
\toprule
\multirow{2}{*}{\textbf{Selection Methods}} & \multicolumn{3}{c}{\textbf{BIRD EX (\%)}} & & \multicolumn{3}{c}{\textbf{Spider EX (\%)}} \\
\cmidrule{2-4} \cmidrule{6-8}
& \textbf{Qwen} & \textbf{DeepSeek} & \textbf{GPT} & & \textbf{Qwen} & \textbf{DeepSeek} & \textbf{GPT} \\
\midrule
\textit{Upper Bound (Pass@5)} & \textit{57.6} & \textit{58.8} & \textit{56.4} & & \textit{84.8} & \textit{77.6} & \textit{76.6} \\
\midrule
\multicolumn{8}{l}{\textbf{\textit{Heuristic-based}}} \\
Random Selection (Baseline) & 37.0 & 48.2 & 46.0 & & 72.3 & 70.9 & 71.9 \\
Execution-Guided Selection~\cite{resd-sql} & 44.4 & 49.8 & 50.2 & & 75.0 & 72.6 & 72.4 \\
Self-Consistency~\cite{alpha-sql} & \underline{46.4} & \underline{51.2} & 49.4 & & \underline{76.5} & 71.6 & 72.2 \\
\midrule
\multicolumn{8}{l}{\textbf{\textit{LLM-based}}} \\
Multiple-Choice Selection~\cite{mcs-sql} & 43.6 & 51.0 & \underline{50.4} & & 74.8 & \underline{72.8} & \underline{72.9} \\
\rowcolor{iceblue} \textbf{Dual-Paradigm Consistency (Ours)} & \textbf{47.6} & \textbf{53.4} & \textbf{51.2} & & \textbf{77.5} & \textbf{73.2} & \textbf{73.3} \\
\bottomrule
\end{tabular}
\vspace{-1em}
\end{table*}

\begin{table}[t!]
\centering
\small
\caption{Performance enhancement of \sys integrated into SOTA Text-to-SQL systems on BIRD ($N=5$).}
\label{tab:workflow_integration}
\vspace{-1em}
\resizebox{\linewidth}{!}{%
\begin{tabular}{l cccc}
\toprule
\multirow{2}{*}{\textbf{Method / Selection}} & \multicolumn{4}{c}{\textbf{Execution Accuracy (\%)}} \\
\cmidrule(l){2-5}
& \textbf{Sim.} & \textbf{Mod.} & \textbf{Cha.} & \textbf{All} \\
\midrule

\multicolumn{5}{l}{\textit{\textbf{Prompting-based}}} \\
\multicolumn{5}{l}{~DAIL-SQL~\cite{dail-sql}} \\ 
~~+ SC & \textbf{64.2} & 46.0 & 32.3 & 48.6 \\
~~+ \sys \textbf{(Ours)} & \cellcolor{iceblue}62.9 & \cellcolor{iceblue}\textbf{49.2} & \cellcolor{iceblue}\textbf{35.3} & \cellcolor{iceblue}\textbf{50.4} \\
\cmidrule[0.5pt](l){2-5} 

\multicolumn{5}{l}{~CHESS~\cite{chess}} \\
~~+ SC & 78.4 & 63.2 & 55.9 & 66.2 \\
~~+ \sys \textbf{(Ours)} & \cellcolor{iceblue}\textbf{78.4} & \cellcolor{iceblue}\textbf{64.8} & \cellcolor{iceblue}\textbf{56.9} & \cellcolor{iceblue}\textbf{67.2} \\

\midrule

\multicolumn{5}{l}{\textit{\textbf{Fine-tuning-based}}} \\
\multicolumn{5}{l}{~OmniSQL-7B~\cite{omnisql}} \\
~~+ SC & 64.8 & 43.6 & 40.2 & 49.2 \\
~~+ \sys \textbf{(Ours)} & \cellcolor{iceblue}\textbf{68.9} & \cellcolor{iceblue}\textbf{44.0} & \cellcolor{iceblue}\textbf{41.2} & \cellcolor{iceblue}\textbf{50.8} \\
\cmidrule[0.5pt](l){2-5}

\multicolumn{5}{l}{~XiYanCoder-7B~\cite{xiyan-sql}} \\
~~+ SC & 70.9 & 49.2 & 36.3 & 53.0 \\
~~+ \sys \textbf{(Ours)} & \cellcolor{iceblue}\textbf{71.6} & \cellcolor{iceblue}\textbf{52.0} & \cellcolor{iceblue}\textbf{40.2} & \cellcolor{iceblue}\textbf{55.4} \\
\bottomrule
\end{tabular}%
} 

\vspace{-1em}
\end{table}

\subsection{Experimental Settings}
\label{sec:exp_settings}

\paragraph{Datasets.} We evaluate \sys on two benchmarks: \textbf{BIRD}~\cite{dataset-bird}, a large-scale dataset focusing on real-world database complexity and external knowledge, and \textbf{Spider}~\cite{dataset-spider}, the standard for evaluating SQL structural generalization. We use the official BIRD Mini-Dev split (500 queries) for efficient agentic reasoning evaluation and the standard Spider Test split (2,147 quries).

\paragraph{Base LLMs.} To assess adaptability, we employ three backbones: \textbf{GPT-5}~\cite{gpt-5} as the frontier proprietary model, and \textbf{DeepSeek-V3.2}~\cite{deepseek} alongside \textbf{Qwen2.5-Coder-7B-Instruct}~\cite{qwen2.5-coder} as state-of-the-art open-source representatives.

\paragraph{Selection Baselines.} We compare \sys against two categories of selection strategies: (1) \textbf{Heuristic-based}: \textit{Random} selection; \textit{Execution-Guided Selection}~\cite{resd-sql}, which selects the first candidate that executes without errors; and \textit{Self-Consistency (SC)}~\cite{alpha-sql}, which selects the SQL yielding the most frequent execution result. (2) \textbf{LLM-based}: \textit{Multiple-Choice Selection (MCS)}~\cite{mcs-sql}, where an LLM is prompted to select the best candidate from the pool.

\paragraph{Integration with Text-to-SQL Systems.} To demonstrate plug-and-play capability, we integrate \sys into: (1) \textbf{Prompting-based} systems, including \textit{DAIL-SQL}~\cite{dail-sql} and \textit{CHESS}~\cite{chess}; and (2) \textbf{Fine-tuning-based} models, including \textit{OmniSQL-7B}~\cite{omnisql} and \textit{XiYanCoder-7B}~\cite{xiyan-sql}.

\paragraph{Evaluation Metrics.} We adopt a multi-dimensional protocol focusing on \textit{Effectiveness} and \textit{Efficiency}. We report \textbf{Execution Accuracy (EX)}~\cite{dataset-bird} as the primary metric, alongside \textbf{Upper Bound (Pass@N)}~\cite{chase-sql} which represents the theoretical maximum performance within the candidate pool. For practical overhead, we measure the average \textbf{Token Cost} and \textbf{Latency} per query to assess the efficiency of the selection process.

\paragraph{Implementation Details.}
For local inference, we deploy \textbf{Qwen2.5-Coder-7B-Instruct}, \textbf{XiYanCoder-7B}, and \textbf{OmniSQL-7B} using the \texttt{vLLM} library~\cite{vllm} on a single NVIDIA A800 (80GB) GPU. Meanwhile, \textbf{DeepSeek-V3.2} and \textbf{GPT-5} are accessed via their official APIs. 
Throughout the \sys process, we set the sampling temperature to $0.7$ and the self-correction limit to $T=3$.

\subsection{Main Results and Analysis}
\label{sec:main_results}


\paragraph{\sys Consistently Outperforms Baselines.}
As shown in Table~\ref{tab:selection_comparison}, \sys achieves state-of-the-art performance across all base LLMs on both BIRD and Spider. 
On the challenging BIRD benchmark, \sys significantly boosts the performance of open-source models; for instance, it improves the EX of \textbf{DeepSeek-V3.2} from 51.2\% (Self-Consistency) to \textbf{53.4\%}. 
Even for the frontier \textbf{GPT-5} model, which already exhibits high baseline accuracy, \sys manages to extract further gains, pushing its EX from 50.4\% (Multiple-Choice Selection) to \textbf{51.2\%}. 
Furthermore, compared to heuristic-based methods like Execution-Guided Selection, \sys demonstrates a superior ability to identify semantically correct SQLs that might otherwise be overlooked by simple execution checks.
We provide a detailed case study comparing \sys with the Self-Consistency baseline in Appendix~\ref{sec:appendix_running_example}.

\paragraph{Versatility in System Integration.}
Table~\ref{tab:workflow_integration} demonstrates \sys integrated into existing SOTA Text-to-SQL systems. 
To ensure a rigorous and cost-effective evaluation, we employ \textbf{Qwen2.5-Coder-7B-Instruct} as the uniform backbone for both the candidate generation in prompting-based systems and the logic execution in the \sys verification agent. 
We observe that \sys serves as a powerful ``plug-and-play'' enhancement layer across all tested systems. 
For prompting-based frameworks, it boosts \textbf{DAIL-SQL} and \textbf{CHESS} by \textbf{+1.8\%} and \textbf{+1.0\%} respectively. 
More importantly, for fine-tuning-based models like \textbf{XiYanCoder-7B}, \sys yields a significant improvement of \textbf{+2.4\%}. 

\begin{table}[t!]
\centering
\small 
\setlength{\tabcolsep}{4pt} 
\caption{Robustness analysis against systematic bias on BIRD using Qwen2.5-Coder-7B-Instruct.}
\label{tab:robustness_bias}
\vspace{-1em}
\begin{tabular}{l ccc}
\toprule
\textbf{Set Category} & \textbf{SC (\%)} & \textbf{DPC (\%)} & \textbf{Gain} \\
\midrule
Overall (Full Set) & 46.4 & 47.6 & +1.2 \\
\midrule
Majority-Correct & \textbf{100.0} & 97.0 & -3.0 \\
Majority-Incorrect & 0.0 & \textbf{21.4} & \textbf{+21.4} \\
\bottomrule
\end{tabular}
\end{table}

\subsection{Robustness Analysis}
\label{sec:robustness}

\paragraph{Resilience to Systematic Bias.}
Traditional SC fails completely (\textbf{0\%} accuracy) on the \textit{Majority-Incorrect Set}, where model-internal biases lead to consistent but erroneous outputs (Table~\ref{tab:robustness_bias}). In contrast, \sys recovers \textbf{21.4\%} of these hard cases by leveraging cross-paradigm verification. Although a minor regression (-3.0\%) occurs in the Majority-Correct set, the significant recovery in biased scenarios proves that \sys serves as a critical safety net when the majority consensus is flawed (see Appendix~\ref{sec:appendix_error} for detailed error analysis).

\paragraph{Scaling with Candidate Size.}

\begin{figure}[t!]
    \centering
    \includegraphics[width=0.48\textwidth]{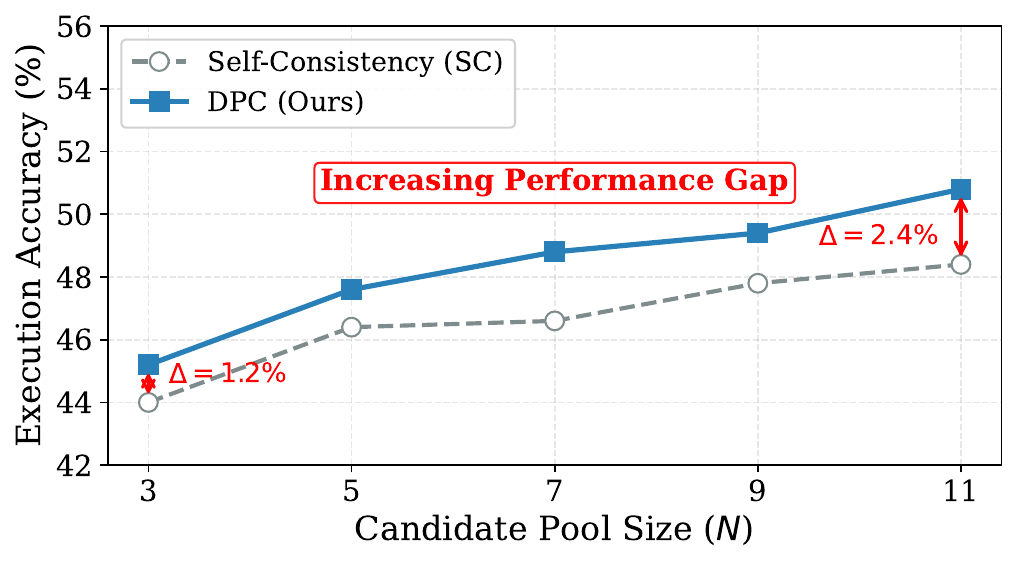}
    \vspace{-2em}
    \caption{Impact of candidate pool size $N$ on BIRD.}
    \label{fig:n_scaling}
    \vspace{-1em}
\end{figure}

As shown in Figure~\ref{fig:n_scaling}, increasing the candidate pool size ($N \in \{3, \dots, 11\}$) widens \sys's lead over SC from \textbf{1.2\%} to \textbf{2.4\%}. This trend validates \sys's superior scalability and proficiency in filtering high-confidence noise within dense search spaces.

\subsection{Efficiency and Cost-Benefit Analysis}
\label{sec:efficiency}

\begin{table}[t!]
\centering
\small
\caption{Efficiency and cost-benefit analysis on BIRD using Qwen2.5-Coder-7B-Instruct. Token cost and latency are reported as average values per question.}
\label{tab:efficiency}
\vspace{-1em}
\resizebox{\columnwidth}{!}{
\begin{tabular}{l ccc}
\toprule
\textbf{Selection Methods} & \textbf{EX (\%)} & \textbf{Tokens (K)} & \textbf{Latency (s)} \\
\midrule
\textit{Heuristic-based} \\
Random Selection & 37.0 & - & $\sim$0.0 \\
Execution-Guided & 44.4 & - & $\sim$0.3 \\
Self-Consistency & 46.4 & - & $\sim$0.8 \\
\midrule
\textit{LLM-based} \\
Multiple-Choice & 43.6 & $\sim$4.2 & $\sim$5.1 \\
\rowcolor{iceblue} \textbf{DPC (Ours)} & \textbf{47.6} & \textbf{$\sim$3.8} & \textbf{$\sim$4.2} \\
\bottomrule
\end{tabular}%
}
\end{table}

\paragraph{Efficiency vs. LLM Baselines.}
Compared to MCS, \sys achieves higher accuracy (\textbf{+4.0\%} EX) with lower overhead, reducing token consumption (4.2k$\to$\textbf{3.8k}) and latency (5.1s$\to$\textbf{4.2s}), as shown in Table~\ref{tab:efficiency}. This efficiency stems from the \slicer Agent, which proactively prunes irrelevant schema, avoiding the redundancy inherent in full-schema selection methods.

\paragraph{Accuracy-Efficiency Trade-off.}
While heuristic baselines (e.g., SC) are near-instantaneous, they compromise precision. \sys outperforms SC by \textbf{1.2\%} on BIRD. In high-stakes domains (e.g., finance) where data integrity is paramount, the marginal latency increase ($\sim$3.6s) is a justifiable investment for the substantial gain in reliability.

\subsection{Ablation Study}

\begin{table}[t]
\centering
\small
\caption{Ablation study of \sys components on the BIRD dataset. The baseline is the full \sys framework equipped with Qwen2.5-Coder-7B-Instruct.}
\label{tab:ablation}
\vspace{-1em}
\setlength{\tabcolsep}{12pt} 
\begin{tabular}{l c c}
\toprule
\textbf{Components} & \textbf{EX (\%)} & \textbf{$\Delta$ EX (\%)} \\
\midrule
\textbf{\sys (Full)} & \textbf{47.6} & \textbf{--} \\
\midrule
w/o Tester Agent & 46.2 & -1.4 \\
w/o Slicer Agent & 46.8 & -0.8 \\
w/o Python Agent & 47.0 & -0.6 \\
w/o BS-F1 Metric & 47.0 & -0.6 \\
w/o Self-Correction & 47.2 & -0.4 \\
\bottomrule
\end{tabular}
\vspace{-1em}
\end{table}

\label{sec:ablation}
Table~\ref{tab:ablation} dissects component-wise contributions on the BIRD set. The \tester Agent proves most critical (\textbf{-1.4\%}), validating the necessity of adversarial feedback for effective MDD construction, followed by the \slicer Agent (\textbf{-0.8\%}) which mitigates schema noise. The remaining modules (\solver Agent, BS-F1, and Self-Correction) collectively contribute to the robustness, confirming the holistic efficacy of the dual-paradigm design.

\section{Related Works}


\paragraph{Text-to-SQL Generation and Selection.}
Recent prompting~\cite{dail-sql, chess, deepvis, nvbench2} and fine-tuning~\cite{xiyan-sql, omnisql} methods achieve strong Pass@K performance on benchmarks such as BIRD and Spider~\cite{nl2sql-survey, data-agent-survey, nl2sql-bugs, deepeye}. 
Yet a critical \textit{Generation-Selection Gap} remains: models can generate correct candidates but often fail to identify them~\cite{agentar-scale-sql}. 
Training-based verifiers~\cite{xiyan-sql, chase-sql} improve selection but require costly annotations and generalize poorly across domains, while training-free methods such as Self-Consistency~\cite{opensearch-sql} and LLM-as-a-Judge~\cite{mcs-sql} are limited by \textit{systematic bias} and \textit{Symbolic Blindness}. 
In contrast, \sys casts selection as deterministic verification in an adversarially synthesized environment.


\paragraph{Data Synthesis for Text-to-SQL.}
To address data scarcity~\cite{DBLP:conf/acl/HuZJLZCLPWHZGDL23, statqa, wu2024chartinsightsevaluatingmultimodallarge, wu2025concisereasoningbiggains, wu2026autowebworldsynthesizinginfiniteverifiable, xie2026visjudgebenchaestheticsqualityassessment}, recent work uses LLMs to synthesize Text-to-SQL corpora. 
Frameworks such as OmniSQL~\cite{omnisql} and SQL-Factory~\cite{sql-factory} emphasize scale through multi-agent generation, while SQLForge~\cite{sql-forge}, SQLord~\cite{sqlord}, and SING-SQL~\cite{sing-sql} improve reliability and domain adaptation via reverse-generation pipelines. 
However, these methods are designed for offline training augmentation and produce static datasets that lack the discriminative validity needed for inference-time verification. 
In contrast, \sys uses synthesis for online verification by dynamically constructing a query-specific Minimal Distinguishing Database (MDD).
\section{Conclusion}
In this paper, we introduce Dual-Paradigm Consistency (\sys), a training-free framework that bridges the generation-selection gap in Text-to-SQL.
By synthesizing an adversarial Minimal Distinguishing Database (MDD) and leveraging the complementary reasoning of SQL and Python, \sys transforms selection from probabilistic guessing into deterministic verification.
Experimental results on BIRD and Spider demonstrate that \sys significantly mitigates systematic bias and achieves state-of-the-art execution accuracy.

\clearpage
\section*{Limitations}
Despite significant gains, \sys incurs higher inference latency than direct generation due to its multi-agent synthesis and execution pipeline. 
Future work will explore \textit{adaptive verification}---triggering DPC only under high uncertainty---to optimize efficiency. 
Additionally, under complex implicit constraints, synthesized environments may structurally diverge from real distributions; thus, improving the \textit{semantic fidelity} of adversarial synthesis remains a critical direction for future exploration.

\section*{Acknowledgments}
This paper was supported by the NSF of China (62402409); Youth S\&T Talent Support Programme of Guangdong Provincial Association for Science and Technology (SKXRC2025461); the Young Talent Support Project of Guangzhou Association for Science and Technology (QT-2025-001); Guangzhou Basic and Applied Basic Research Foundation (2026A1515010269, 2025A04J3935, 2023A1515110545); and Guangzhou-HKUST(GZ) Joint Funding Program (2025A03J3714).

\bibliography{main}

\clearpage
\newpage
\appendix

\section{Prompts for \sys Agents}
\label{sec:appendix_prompts}

To facilitate reproducibility and provide transparency into the \sys framework, we present the detailed prompt templates used for the three core agents: \slicer, \tester, and \solver. 
Each agent operates with a distinct system prompt that defines its specific role, reasoning instructions, and strict output formats (e.g., JSON or Python code), ensuring robust communication and error handling within the multi-agent pipeline.

\subsection{Prompts for \slicer Agent}
The \slicer Agent performs \textit{schema pruning} to extract a Minimal Schema Slice, identifying only the tables and columns strictly necessary for the subsequent verification pipeline.

\begin{promptframe}{System Prompt: \slicer Agent}
\begin{lstlisting}[style=promptstyle]
You are a database expert. Your task is to identify the minimum, non-duplicate set of tables and columns required to execute the provided Candidate SQL Queries.

You should:
1. Analyze the provided Candidate SQL Queries.
2. Identify all tables and columns from the Full Database Schema that are actually used in these SQLs (SELECT, JOIN, WHERE, GROUP BY, etc.).
3. Ensure you only use table and column names exactly as they appear in the Full Database Schema.
4. Provide a concise thinking process before the final result.

Your output MUST follow this format:
<thinking>
[Your step-by-step analysis here]
</thinking>
<result>
{
    "relevant_schema": [
        {
            "table": "table_name",
            "columns": ["column1", "column2"]
        },
        ...
    ]
}
</result>
IMPORTANT: The content inside <result> MUST be a valid, standard JSON string that can be parsed by `json.loads()`. DO NOT include any comments (like // or /* */) or extra text inside the JSON block.
\end{lstlisting}
\end{promptframe}

\begin{promptframe}{User Prompt Template: \slicer Agent}
\begin{lstlisting}[style=promptstyle]
Full Database Schema:
(*@\pvar{\{full\_schema\}}@*)

Candidate SQL Queries:
(*@\pvar{\{candidate\_sqls\}}@*)

Please identify the relevant tables and columns used in the Candidate SQL Queries, following the output format defined in the system prompt.
\end{lstlisting}
\end{promptframe}

\begin{promptframe}{Retry Prompt Template: \slicer Agent}
\begin{lstlisting}[style=promptstyle]
The previously identified Schema Slice has issues. Error details:
---
(*@\pvar{\{error\_message\}}@*)
---

Please analyze the error (it could be a format issue, missing tables/columns, or incorrect names) and provide the corrected, complete Schema Slice. 
Ensure your response strictly follows the output format defined in the system prompt (including <thinking> and <result> tags).
The content inside <result> MUST be a valid, standard JSON string without any comments.
\end{lstlisting}
\end{promptframe}

\subsection{Prompts for \tester Agent}
The \tester Agent is responsible for generating adversarial data (Minimal Distinguishing Database) to expose logical discrepancies.

\begin{promptframe}{System Prompt: \tester Agent}
\begin{lstlisting}[style=promptstyle]
You are a database QA engineer. Your task is to generate a minimal set of synthetic test data (rows) that will cause two different SQL queries to return DIFFERENT results.

You should:
1. Analyze the Natural Language Question and the Candidate SQLs.
2. Identify the logical difference between SQL 1 and SQL 2 (e.g., a filter condition, a join type, or an aggregation).
3. Generate a sufficient but minimal set of data that specifically triggers this logical difference.
4. Ensure the data adheres to the Sliced Database Schema (correct table/column names, types, and foreign key relationships).
5. Leverage metadata in the Schema: Use column descriptions, value descriptions, and example values provided in the schema to ensure the generated test data is realistic and follows the expected data distribution/format of the original database.
6. Provide a concise thinking process before the final result.

Your output MUST follow this format:
<thinking>
[Your analysis of why the SQLs differ and how your data will expose that]
</thinking>
<result>
{
    "test_data": {
        "table_name1": [
            {"column1": value1, "column2": value2},
            ...
        ],
        "table_name2": [...]
    }
}
</result>
IMPORTANT: The content inside <result> MUST be a valid, standard JSON string that can be parsed by `json.loads()`. DO NOT include any comments (like // or /* */) or extra text inside the JSON block.
\end{lstlisting}
\end{promptframe}

\begin{promptframe}{User Prompt Template: \tester Agent}
\begin{lstlisting}[style=promptstyle]
Sliced Database Schema:
(*@\pvar{\{sliced\_schema\}}@*)

Natural Language Question: (*@\pvar{\{question\}}@*)
(*@\pvar{\{evidence\_str\}}@*)

Candidate SQL 1 (Champion):
(*@\pvar{\{sql\_1\}}@*)

Candidate SQL 2 (Challenger):
(*@\pvar{\{sql\_2\}}@*)

Please generate the test data that makes SQL 1 and SQL 2 yield different results, following the output format defined in the system prompt.
\end{lstlisting}
\end{promptframe}

\begin{promptframe}{Retry Prompt Template: \tester Agent}
\begin{lstlisting}[style=promptstyle]
The previously generated test data has issues or is INEFFECTIVE. Error details:
---
(*@\pvar{\{error\_message\}}@*)
---

Please analyze the logic difference between SQL 1 and SQL 2 again, and provide a corrected set of test data that yields DIFFERENT results. 
Ensure your response strictly follows the output format defined in the system prompt (including <thinking> and <result> tags).
The content inside <result> MUST be a valid, standard JSON string without any comments.
\end{lstlisting}
\end{promptframe}

\subsection{Prompts for \solver Agent}
The \solver Agent functions as the reference anchor, executing Python logic on the synthesized MDD to determine the correct result.

\begin{promptframe}{System Prompt: \solver Agent}
\begin{lstlisting}[style=promptstyle]
You are a Python data scientist expert in Pandas. Your task is to write a Python script to answer a natural language question based on provided database tables (loaded as Pandas DataFrames).

You should:
1. Analyze the schema and the provided test data.
2. Use the provided DataFrames (already available in the namespace with their table names).
3. Write clean, efficient Pandas code to compute the answer.
4. IMPORTANT: Store the final result in a variable named 'result'. 
5. The 'result' MUST ALWAYS be a pandas DataFrame. Even for single values or lists, wrap them in a DataFrame.
6. COLUMN SELECTION: The final DataFrame MUST ONLY contain columns that are explicitly asked for in the question. Do not include extra or redundant columns.
7. COLUMN ORDERING: The order of columns in the final DataFrame MUST strictly follow the order mentioned in the natural language question.

Your output MUST follow this format:
<thinking>
[Your step-by-step logic for solving the problem using Pandas]
</thinking>
<result>
[Your Python code here]
</result>
\end{lstlisting}
\end{promptframe}

\begin{promptframe}{User Prompt Template: \solver Agent}
\begin{lstlisting}[style=promptstyle]
(*@\pvar{\{test\_data\_with\_types\}}@*)

### Database Relationships (PK/FK):
(*@\pvar{\{relationships\}}@*)

### Available DataFrames (Pandas Variables):
(*@\pvar{\{df\_names\}}@*)

### Natural Language Question:
(*@\pvar{\{question\}}@*)
(*@\pvar{\{evidence\_str\}}@*)

Please write the Pandas code to solve the question. 
Remember:
- Only return columns explicitly asked for in the question.
- The column order must match the question's order.
- Ensure the final answer is stored in the `result` variable as a DataFrame.
\end{lstlisting}
\end{promptframe}

\begin{promptframe}{Retry Prompt Template: \solver Agent}
\begin{lstlisting}[style=promptstyle]
The previously generated code has issues. Error details:
---
(*@\pvar{\{error\_message\}}@*)
---

Please analyze the error and provide the corrected Python code. 
Ensure your response strictly follows the output format defined in the system prompt (including <thinking> and <result> tags).
\end{lstlisting}
\end{promptframe}

\section{Case Study: Identifying Missing Constraints}
\label{sec:appendix_running_example}

We present a detailed case study (QID: 1500, Source: \textit{BIRD Mini-Dev Split}) to demonstrate how \sys leverages dual-paradigm execution to detect missing logical constraints (e.g., missing JOINs) that heuristic methods often overlook.

\subsection{The Conflict: Missing Table Constraint}
\textbf{Question:} \textit{``Please list the product description of the products consumed in September, 2013.''}

The baseline SC method selects a \textbf{Champion SQL} that relies solely on the \texttt{transactions\_1k} table, assuming the transaction date is sufficient. 
In contrast, \sys selects a \textbf{Challenger SQL} that introduces an additional JOIN with the \texttt{yearmonth} table, implying that a valid consumption record requires a corresponding monthly entry.

\begin{tcolorbox}[enhanced, title=\textbf{Step 1: Logical Divergence}, boxrule=0.5pt, colback=white, colframe=gray!50, coltitle=black]
    \begin{minipage}[t]{0.48\linewidth}
        \raggedright 
        \textbf{\textcolor{red}{Champion SQL}} \\ 
        \textbf{\textcolor{red}{(Incorrect):}} 
        
        \begin{lstlisting}[language=SQL, basicstyle=\scriptsize\ttfamily, breaklines=true, keywordstyle=\color{blue}]
SELECT p.Description
FROM transactions_1k t
JOIN products p ON ...
WHERE t.Date LIKE '201309%'
        \end{lstlisting}
        \vspace{-0.5em}
        \small{\textit{$\to$ MISSING constraint: `yearmonth' table.}}
    \end{minipage}
    \hfill
    \begin{minipage}[t]{0.48\linewidth}
        \raggedright 
        \textbf{\textcolor{green!60!black}{Challenger SQL}} \\ 
        \textbf{\textcolor{green!60!black}{(Correct):}} 
        
        \begin{lstlisting}[language=SQL, basicstyle=\scriptsize\ttfamily, breaklines=true, keywordstyle=\color{blue}]
SELECT p.Description
FROM products p
JOIN transactions_1k t ...
JOIN yearmonth ym ON ...  
WHERE ym.Date LIKE '201309%'
        \end{lstlisting}
        \vspace{-0.5em}
        \small{\textit{$\to$ ENFORCES constraint: `yearmonth' table.}}
    \end{minipage}
\end{tcolorbox}

\subsection{Adversarial Synthesis (MDD)}
The \tester agent identifies that the Champion SQL is a \textit{superset} of the Challenger. To expose the error, it constructs a ``trap'' scenario in the \textit{Minimal Distinguishing Database} (MDD):
\begin{enumerate}
    \item It creates a valid transaction for \textbf{Customer 100} in September 2013 (Product: `LPG').
    \item Crucially, it \textbf{omits} Customer 100 from the \texttt{yearmonth} table for that period.
\end{enumerate}

This data distribution creates a decisive split:
\begin{itemize}
    \item \textbf{Champion Result:} Returns \texttt{[`LPG']} (Matches transaction date).
    \item \textbf{Challenger Result:} Returns \texttt{Empty} (Fails INNER JOIN with \texttt{yearmonth}).
\end{itemize}

\begin{table}[h]
    \centering
    \small
    \caption{\textbf{Synthesized MDD.} The agent creates a ``Ghost Transaction'' (Row 1) that lacks a parent record in the YearMonth table (Bottom).}
    \label{tab:mdd_debit}
    
    \begin{tabular}{c c c l}
        \multicolumn{4}{c}{\textbf{Table: transactions\_1k}} \\ 
        \toprule
        \textbf{TxID} & \textbf{CustID} & \textbf{Date} & \textbf{Product} \\
        \midrule
        \rowcolor{green!10} 1 & \textbf{100} & 20130915 & \textbf{LPG (ID:2)} \\
        \bottomrule
    \end{tabular}
    
    \vspace{1em} 
    
    \begin{tabular}{c c}
        \multicolumn{2}{c}{\textbf{Table: yearmonth}} \\ 
        \toprule
        \textbf{CustID} & \textbf{Date} \\
        \midrule
        200 & 201309 \\
        300 & 201309 \\
        \rowcolor{red!10} \multicolumn{2}{c}{\textit{-- Cust 100 is Missing! --}} \\
        \bottomrule
    \end{tabular}
\end{table}

\subsection{Dual-Paradigm Verification}
The \solver agent executes the verification logic using Pandas. As shown below, the Python agent explicitly merges the \texttt{yearmonth} dataframe, validating the structural necessity of this table.

\begin{tcolorbox}[enhanced, title=\textbf{Step 3: Python Logic Alignment}, boxrule=0.5pt, colback=gray!5, colframe=gray!60, coltitle=black, drop shadow]
\begin{lstlisting}[language=Python, basicstyle=\scriptsize\ttfamily, keywordstyle=\color{blue}, commentstyle=\color{green!50!black}, breaklines=true]
# Solver Agent's Logic:
# 1. Filter yearmonth table first. 
# Note: Python explicitly uses the yearmonth constraint.
mask = yearmonth['Date'].str.startswith('201309')
filtered_ym = yearmonth[mask]

# 2. explicit MERGE acts as a filter (Inner Join)
# This line kills the result because Customer 100 is NOT in filtered_ym
merged_tx = pd.merge(filtered_ym, transactions_1k, on='CustomerID')

result = pd.merge(merged_tx, products, ...)[['Description']]
\end{lstlisting}
\end{tcolorbox}

\paragraph{Selection Result:} 
The Python execution returns an \textbf{Empty DataFrame}, perfectly aligning with the Challenger SQL's result. 
\sys determines that the Champion SQL hallucinated a simplification (ignoring the \texttt{yearmonth} table) that contradicts the cross-paradigm consensus. 
Thus, \sys correctly identifies the missing constraint and selects the Challenger SQL.
\section{Error Analysis}
\label{sec:appendix_error}

To deliver a thorough analysis, we present a systematic error analysis comparing our \sys framework against the \textbf{Self-Consistency (SC)} baseline, based on execution results from the BIRD Mini-Dev split. We first introduce our error taxonomy used throughout the analysis, then provide a statistical overview of error disagreements, followed by detailed case studies illustrating how \sys’s dual-paradigm verification resolves or fails to resolve specific error types.

\subsection{Error Taxonomy}
We adopt a two-level error taxonomy to categorize discrepancies between \sys and SC, adapted from prior studies on Text-to-SQL errors. Errors are classified into two primary categories:

\begin{enumerate}
    \item \textbf{Semantic-Level Errors (A)}  
    Arise from misunderstandings of query semantics, such as incorrect filtering, aggregation, ordering, or result representation. These errors often stem from natural language ambiguity or misinterpretation of query intent.

    \item \textbf{Structural/Schema-Level Errors (B)}  
    Involve incorrect schema linking, join path selection, or table/column references. These errors reflect failures in mapping the question to the underlying database structure.
\end{enumerate}

A detailed breakdown of error subcategories, along with representative examples, is provided in Appendix Table \ref{tab:error-taxonomy}. Based on our validation on the BIRD Mini-Dev split, among the cases where \sys succeeds while SC fails, three error subtypes dominate the distribution, as follows:
\textbf{Result Representation Error (A5)} constitutes \textbf{41.7\%} of the observed failures, followed by \textbf{Schema Linking Error (B3)} at \textbf{25\%}, and \textbf{Predicate Misalignment Filter (A1)} accounting for \textbf{16.7\%}.

\begin{table*}[t]
\centering
\small
\setlength{\tabcolsep}{6pt}
\caption{Error Taxonomy for \sys vs.\ SC Comparison}
\label{tab:error-taxonomy}

\begin{tabular}{c c l p{8cm}}
\toprule
\textbf{Category} & \textbf{Code} & \textbf{Error Type} & \textbf{Description} \\
\midrule

\multirow{5}{*}{\textbf{Semantic (A)}} 
& A1 & Predicate Misalignment 
& Incorrect filtering logic, value hallucination, misuse of operators (e.g., \texttt{LIKE}, \texttt{IN}, \texttt{BETWEEN}), or logical connectors (\texttt{AND}/\texttt{OR}). \\

& A2 & Aggregation Misuse 
& Misapplication of aggregate functions (\texttt{COUNT}, \texttt{SUM}, \texttt{AVG}, etc.). \\

& A3 & Grouping Error 
& Missing or incorrect \texttt{GROUP BY} columns. \\

& A4 & Ordering \& Superlative 
& Incorrect \texttt{ORDER BY} or \texttt{LIMIT} usage, especially in queries involving ``most,'' ``least,'' or ranking. \\

& A5 & Result Format Error 
& Issues in the \texttt{SELECT} clause, including wrong data types, missing or extra columns, and formatting errors (e.g., rounding, casting). \\

\midrule

\multirow{3}{*}{\textbf{Structural (B)}} 
& B1 & Wrong Join Path 
& Incorrect table joins, including missing, extra, or incorrect tables. \\

& B2 & Join Condition Error 
& Incorrect \texttt{ON} conditions in join operations. \\

& B3 & Schema Linking Error 
& Correct semantic understanding but failure in composing multi-table logic, often due to incorrect column or table mapping. \\

\bottomrule
\end{tabular}
\end{table*}

\subsection{Case Studies: Semantic-Level Errors}

\subsubsection{Predicate Misalignment}

\begin{tcolorbox}[
    enhanced,
    title=\textbf{(QID: 1464 from BIRD Mini-Dev Split)},
    boxrule=0.5pt,
    colback=white,
    colframe=gray!50,
    coltitle=black
]

\begin{minipage}[t]{0.48\linewidth}
    \raggedright
    \textbf{\textcolor{red}{SC SQL}} \\
    \textbf{\textcolor{red}{(Incorrect):}}
    
    \begin{lstlisting}[language=SQL, basicstyle=\scriptsize\ttfamily, breaklines=true, keywordstyle=\color{blue}]
SELECT m.first_name, m.last_name, i.amount
FROM member AS m
JOIN income AS i ON m.member_id = i.link_to_member
WHERE i.date_received = '9/9/2019';
    \end{lstlisting}
    \vspace{-0.5em}
    \small{\textit{$\to$ INCORRECT predicate value: uses colloquial date format not supported by the database schema.}}
\end{minipage}
\hfill
\begin{minipage}[t]{0.48\linewidth}
    \raggedright
    \textbf{\textcolor{green!60!black}{\sys SQL}} \\
    \textbf{\textcolor{green!60!black}{(Correct):}}
    
    \begin{lstlisting}[language=SQL, basicstyle=\scriptsize\ttfamily, breaklines=true, keywordstyle=\color{blue}]
SELECT T2.first_name, T2.last_name, T1.amount
FROM income AS T1
JOIN member AS T2 ON T1.link_to_member = T2.member_id
WHERE STRFTIME('%Y-%m-%d', T1.date_received) = '2019-09-09';
    \end{lstlisting}
    \vspace{-0.5em}
    \small{\textit{$\to$ CORRECT predicate grounding: enforces ISO date format consistent with schema constraints.}}
\end{minipage}

\end{tcolorbox}


\paragraph{Predicate Misalignment (Value Hallucination).}  
Case ID~1464 illustrates a representative predicate-level error where SC relies on natural language priors to generate a colloquial date literal (\texttt{'9/9/2019'}) that does not conform to the database’s expected format, resulting in execution failure despite correct intent detection.

\paragraph{Schema-Grounded Correction.}  
\sys successfully aligns the predicate value with schema constraints by explicitly normalizing the date representation to the ISO format (\texttt{'2019-09-09'}), ensuring compatibility with the underlying database.

\paragraph{Robust Literal Alignment.}  
This case demonstrates that effective text-to-SQL generation requires grounding literal value predictions in database-specific formatting rules rather than surface-level textual patterns, highlighting \sys’s advantage over SC in handling \textbf{A1 Predicate Misalignment} errors.

\subsubsection{Result Representation Error}

\begin{tcolorbox}[
    enhanced,
    title=\textbf{(QID: 227 from BIRD Mini-Dev Split)},
    boxrule=0.5pt,
    colback=white,
    colframe=gray!50,
    coltitle=black
]

\begin{minipage}[t]{0.48\linewidth}
    \raggedright
    \textbf{\textcolor{red}{SC SQL}} \\
    \textbf{\textcolor{red}{(Incorrect):}}
    
    \begin{lstlisting}[language=SQL, basicstyle=\scriptsize\ttfamily, breaklines=true, keywordstyle=\color{blue}]
SELECT CAST(SUM(CASE WHEN label = '+' THEN 1 ELSE 0 END) AS REAL)
       * 100 / COUNT(*) AS percent
FROM molecule;
    \end{lstlisting}
    \vspace{-0.5em}
    \small{\textit{$\to$ INCORRECT result representation: missing \texttt{ROUND(..., 3)}, does not satisfy the required precision.}}
\end{minipage}
\hfill
\begin{minipage}[t]{0.48\linewidth}
    \raggedright
    \textbf{\textcolor{green!60!black}{\sys SQL}} \\
    \textbf{\textcolor{green!60!black}{(Correct):}}
    
    \begin{lstlisting}[language=SQL, basicstyle=\scriptsize\ttfamily, breaklines=true, keywordstyle=\color{blue}]
SELECT ROUND(
    CAST(SUM(CASE WHEN label = '+' THEN 1 ELSE 0 END) AS REAL)
    * 100 / COUNT(molecule_id),
    3
) AS percent
FROM molecule;
    \end{lstlisting}
    \vspace{-0.5em}
    \small{\textit{$\to$ CORRECT result representation: explicit casting and rounding ensure alignment with the question.}}
\end{minipage}

\end{tcolorbox}


\paragraph{Result Representation Awareness.}  
In this case, SC correctly captures the underlying computation for estimating the proportion of carcinogenic molecules, but fails to align with the question’s output specification by omitting explicit precision control (i.e., \texttt{ROUND(..., 3)}).

\paragraph{Question-Aligned Refinement.}  
\sys refines the \texttt{SELECT} clause by enforcing both numeric type casting and result formatting, ensuring that the returned value strictly adheres to the requirement of three decimal places.

\paragraph{Consistency Selection.}  
By favoring SQL candidates whose execution results match the question-defined output format, \sys successfully corrects result-level mismatches on top of a logically correct foundation. This demonstrates its advantage over SC in handling \textbf{A5 Result Representation Errors}.

\subsection{Case Studies: Structural/Schema-Level Errors}

\subsubsection{Wrong Join Path}
\begin{tcolorbox}[
    enhanced,
    title=\textbf{(QID: 1340 from BIRD Mini-Dev Split)},
    boxrule=0.5pt,
    colback=white,
    colframe=gray!50,
    coltitle=black
]

\begin{minipage}[t]{0.48\linewidth}
    \raggedright
    \textbf{\textcolor{red}{SC SQL}} \\
    \textbf{\textcolor{red}{(Incorrect):}}
    
    \begin{lstlisting}[language=SQL, basicstyle=\scriptsize\ttfamily, breaklines=true, keywordstyle=\color{blue}]
SELECT 
SUM(CASE WHEN SUBSTR(T1.event_date, 1, 4)
= '2019' THEN T2.spent ELSE 0 END) -
SUM(CASE WHEN SUBSTR(T1.event_date, 1, 4)
= '2020' THEN T2.spent ELSE 0 END) AS difference
FROM event AS T1
INNER JOIN 
budget AS T2 ON T1.event_id = T2.link_to_event
INNER JOIN 
expense AS T3 ON T2.budget_id = T3.link_to_budget
WHERE T1.status = 'Student_Club'

    \end{lstlisting}
    \vspace{-0.5em}
    \small{\textit{$\to$Incorrect Join Path: Utilized unnecessary extra tables for redundant filtering.}}
\end{minipage}
\hfill
\begin{minipage}[t]{0.48\linewidth}
    \raggedright
    \textbf{\textcolor{green!60!black}{Challenger SQL}} \\
    \textbf{\textcolor{green!60!black}{(Correct):}}
    
    \begin{lstlisting}[language=SQL, basicstyle=\scriptsize\ttfamily, breaklines=true, keywordstyle=\color{blue}]
SELECT SUM(CASE WHEN
SUBSTR(T1.event_date, 1, 4)
= '2019' THEN T2.spent ELSE 0 END)
- SUM(CASE WHEN
SUBSTR(T1.event_date, 1, 4)
= '2020' THEN T2.spent ELSE 0 END)
AS difference
FROM event AS T1
JOIN budget AS T2
ON T1.event_id = T2.link_to_event

    \end{lstlisting}
    \vspace{-0.5em}
    \small{\textit{$\to$ CORRECT Join Path: Understand the natural language query and identifying the appropriate tables.}}
\end{minipage}

\end{tcolorbox}

\paragraph{Adversarial Environment Synthesis}
\sys constructs a Minimal Distinguishing Database (MDD) tailored to expose logical discrepancies between candidate queries. The MDD is engineered to be small enough to fit within the model’s context window yet rich enough to differentiate between semantically distinct SQLs. This allows \sys to execute both candidates in a fully observable, controlled environment.

\paragraph{Dual-Paradigm Verification}
Instead of relying solely on SQL execution or internal model priors, \sys leverages a Python/Pandas reference generated by the Solver agent. The Python solution, derived from the same MDD, serves as a high-confidence proxy ground truth. By comparing the execution results of the SQL candidates against this independent imperative reference, \sys can objectively determine which SQL aligns with the correct logic.

\paragraph{Avoidance of Systematic Bias}
Unlike SC, which tends to reinforce the model’s internal biases, \sys’s cross-paradigm verification breaks the cycle of self-confirmation. By grounding the decision in explicit execution evidence from an adversarial environment and an independent reasoning path, \sys avoids falling into consensus traps on erroneous logic.

In this case, \sys correctly selects the simpler and accurate query that joins only event and budget tables, avoiding the unnecessary and erroneous inclusion of the expense table and the extraneous status filter.

\subsubsection{Schema Linking Error}

\begin{tcolorbox}[
    enhanced,
    title=\textbf{(QID: 1134 from BIRD Mini-Dev Split)},
    boxrule=0.5pt,
    colback=white,
    colframe=gray!50,
    coltitle=black
]

\begin{minipage}[t]{0.48\linewidth}
    \raggedright
    \textbf{\textcolor{red}{SC SQL}} \\
    \textbf{\textcolor{red}{(Incorrect):}}
    
    \begin{lstlisting}[language=SQL, basicstyle=\scriptsize\ttfamily, breaklines=true, keywordstyle=\color{blue}]
SELECT (SELECT jumping
FROM Player_Attributes
WHERE player_api_id = 6) -
(SELECT jumping
FROM Player_Attributes
WHERE player_api_id = 23) AS difference;
    \end{lstlisting}
    \vspace{-0.5em}
    \small{\textit{$\to$ INCORRECT column mapping: uses \texttt{player\_api\_id} instead of \texttt{id}.}}
\end{minipage}
\hfill
\begin{minipage}[t]{0.48\linewidth}
    \raggedright
    \textbf{\textcolor{green!60!black}{Challenger SQL}} \\
    \textbf{\textcolor{green!60!black}{(Correct):}}
    
    \begin{lstlisting}[language=SQL, basicstyle=\scriptsize\ttfamily, breaklines=true, keywordstyle=\color{blue}]
SELECT (SELECT jumping
FROM Player_Attributes
WHERE id = 6) -
(SELECT jumping
FROM Player_Attributes
WHERE id = 23) AS difference;
    \end{lstlisting}
    \vspace{-0.5em}
    \small{\textit{$\to$ CORRECT column mapping: uses \texttt{id} as player identifier.}}
\end{minipage}

\end{tcolorbox}

\textbf{Constructing Minimum Differentiating Database (MDD):}  
The \textsc{Slicer} and \textsc{Tester} of \sys collaborate to synthesize a minimal execution environment in which the two candidate column mappings (\texttt{player\_api\_id} vs.\ \texttt{id}) produce different results, thereby exposing the underlying semantic discrepancy.

\textbf{Dual Paradigm Verification}:  
\sys reimplements the SQL logic using Python scripts on the MDD, producing a high-confidence reference output that serves as an execution-level oracle.

\textbf{Consistency Alignment}:  
By comparing SQL execution results against the Python reference using the \textbf{BS-F1} metric, \sys selects the semantically consistent and correct SQL formulation (i.e., using \texttt{id} instead of \texttt{player\_api\_id}).

Therefore, \sys successfully overcomes the symbolic blind spot and system bias inherent in SC, achieving precise alignment with user intent.

\subsection{Discussion: How \sys Addresses These Errors}
\sys’s dual-paradigm design systematically targets two fundamental error classes prevalent in SC and other training-free methods:

\begin{enumerate}
    \item \textbf{Schema Mapping Errors (Category B3)}  
    LLMs suffer from \emph{partial observability}—they only see a schema snippet during inference. \sys’s \slicer agent extracts a relevant schema subgraph and validates it via \emph{dry-run execution} on an empty database. If joins or column references fail, the error feedback iteratively corrects the schema linking before any data synthesis. This acts as \emph{schema-level unit testing}, eliminating B3 errors before execution.

    \item \textbf{Result Representation Errors (Category A5)}  
    LLMs struggle to infer exact output format (columns, types, rounding). \sys’s \solver generates a parallel Python/Pandas script on the same micro-database, providing a high-confidence reference output \(E_{py}\). The \textbf{BS-F1} metric then compares SQL results against \(E_{py}\), penalizing formatting mismatches, extra/missing columns, or incorrect data types. This cross-paradigm consistency check enforces output alignment.
\end{enumerate}

SC’s failures often stem from \emph{systematic bias} and lack of execution grounding. \sys mitigates these by enforcing schema validation and cross-paradigm consistency, exposing errors through adversarial database construction and deterministic matching.

\end{document}